\documentclass[aps,a4paper,showpacs,nofootinbib]{revtex4}
\usepackage{graphicx}
\usepackage{color}
\usepackage{amsmath}
\usepackage{amssymb}
\usepackage{enumerate}
\usepackage{graphics,epsfig,subfigure}
\usepackage{epstopdf}
\usepackage{color}

\newcommand\be{\begin{equation}}
\newcommand\ee{\end{equation}}
\newcommand\beq{\begin{equation}}
\newcommand\eeq{\end{equation}}
\newcommand\ba{\begin{eqnarray}}
\newcommand\ea{\end{eqnarray}}

\newcommand\pt{\partial}

\begin{document}

\title{De Sitter braneworld and gravitational waves}
\author{Dong-Yu Li$^{1}$\footnote{lidongyu@email.ncu.edu.cn}, Zhao-Xiang Wu$^{2}$  \footnote{wuzhaoxiang@email.ncu.edu.cn}, Hao Hu$^{1}$\footnote{huhao@email.ncu.edu.cn}, and Bao-Min Gu$^{1,3}$\footnote{gubm@ncu.edu.cn, corresponding author}}

\affiliation{$^1$Department of Physics, Nanchang University,
Nanchang 330031, China}
\affiliation{$^2$Department of Qianhu, Nanchang University,
Nanchang 330031, China}
\affiliation{$^3$ Center for Relativistic Astrophysics and High Energy Physics, Nanchang University, Nanchang, 330031, China}

\begin{abstract}
We study the braneworld theory constructed by multi scalar fields. The model contains a smooth and infinitely large extra dimension, allowing the background fields propagating in it. We give a de Sitter solution for the four-dimensional cosmology as a good approximation to the early universe inflation. We show that the graviton has a localizable massless mode, and a series of continuous massive modes, separated by a mass gap. There could be a normalizable massive mode, depending on the background solution. The gravitational waves of massless mode evolve the same as the four dimensional theory, while that of the massive modes evolve greatly different from the massless mode.
\end{abstract}

\maketitle



\section{Introduction}
Extra dimension theories provide intriguing ways to address the fundamental problems like the gauge hierarchy problem \cite{Arkani-Hamed1998a,Antoniadis1998a,Randall1999},
the cosmological constant problem \cite{Arkani-Hamed1999a,Cline:1999ts,Dvali:2000xg,Chen:2000at,Forste:2000ft,Navarro:2003vw}, the mass hierarchy of fermions \cite{Gherghetta:2000qt,Huber:2000ie,Dvali:2002pe}, etc. The conventional idea that the extra dimension should be compactified to very small size was also challenged, since even infinitely large extra dimension could be consistent with the observed four dimensional world \cite{Randall1999a}. These progresses extended the physics beyond the standard model of both particle physics and cosmology, hence the extra dimension could be a fundamental element of our world.

An important prediction of extra dimension theories is the Kaluza-Klein graviton. Due to the extra dimension, the graviton generally have a series of Kaluza-Klein modes, with different mass spectrum depending on the model setup. For example, in large extra dimension theory \cite{Arkani-Hamed1998a,Antoniadis1998a}, the typical mass of gravitons could be $\sim 10^{-3} \mathrm{eV}$, originating from the extra dimension size, which is sub-millimetre for two extra dimensions. In the warped extra dimension theory \cite{Randall1999}, our world is confined on a 3-brane (IR brane) embedded in a higher dimensional bulk. By introducing the UV brane, the radius of the extra dimension is of an order larger than the Planck length, leading to a TeV size mass gap of Kaluza-Klein gravitons. These particles would give rise to new phenomenon beyond the standard model that could be detected by high energy accelerators \cite{KumarRai:2003kk,CMS:2012yf,ATLAS:2014wsp}.

Apart from particle physics, extra dimension theories also alter the cosmological evolution of our universe. For example, in Randall-Sundrum  model \cite{Randall1999a} the Einstein equation has a modification compared with the that of general relativity in four dimensions\cite{Shiromizu2000}. This modification turns out to be dominant at high energy scales and can be neglected at low energy scales. This implies that the effects of extra dimension would be significant in early universe such as inflation \cite{Maartens2004a}. In contrast, in large extra dimension theory the new physics appear at low energy scales ($\sim 10^{-3}$ eV), hence it is reasonable to consider the effects of extra dimension when studying the phenomena of late time universe in this model. Generally speaking, it is difficult to solve a five dimensional spacetime background unless some symmetries are imposed, for some examples see References \cite{Binetruy:1999ut,Csaki:1999mp,Deffayet:2000uy,Langlois:2000ns,Quevedo:2002xw,Langlois:2002bb,
Brax:2003fv,Brax:2004xh}. In this work we consider an infinitely large extra dimension theory. The extra dimension is warped because of the scalar fields. In our model the matter fields are not confined on a 3-brane but have components along the fifth dimension. The matter fields should have some special localization mechanisms \cite{Ringeval:2001cq,Dubovsky:2001pe,Liu:2008wd,Liu:2009ve} so that the model is consistent with observations. We consider a de Sitter expansion as an approximation to the early universe inflation. In such a theory, what we are most interested in is the gravitational waves.
The paper is organized as follows. In section \ref{section II} we will give the model setup and the background solutions. In section \ref{section III} we will study the gravitational waves. We will give the mass spectrum of the Kaluza-Klein gravitons and analyze the evolution of gravitational waves. Finally, the conclusions are given in section \ref{concs}.

\section{The model}\label{section II}
We consider the background described by the following line element,
\beq
\mathrm{d}s_5^2=b^2(y)\left(-\mathrm{d}t^2+a^2(t)\delta_{ij}\mathrm{d}x^i \mathrm{d}x^j\right)+\mathrm{d}y^2,
\label{metricassumption}
\eeq
where $b(y)$ is the warp factor characterizing the curved extra dimension, and $a(t)$ is the usual scale factor.
The action of the theory is given by
\beq
S=\int \mathrm{d}^5x \sqrt{-g}\left(\frac{R}{2\kappa_5}
-\frac{1}{2}\pt^\mathrm{P}\Phi_I\pt_\mathrm{P}\Phi^I - V\left(\Phi^I\right)\right).
\eeq
Here $\Phi_I=\mathcal{G}_{IJ}\Phi^I$ with $I=1,2,3...$ representing the field index, and $\mathcal{G}_{IJ}$ is the metric in field space.
Note that in this work the letters, i.e. $P, Q, M, N...$ represent the five-dimensional coordinates, Greek letters for the four-dimensional, and lowercase for the three-dimensional. Generally, the field space could be warped, leading to non-canonical scalar fields and couplings between different fields. In this work we consider a flat field space for simplicity, i.e. $\mathcal{G}_{IJ}=\eta_{IJ}$. Under this assumption, the field equation for the scalar is
\be
\Phi_I''+4\frac{b'}{b}\Phi_I'+\frac{\partial V}{\partial\Phi^I}=0.
\ee
The Einstein equations are
\ba
&&00: 3\frac{b'^2}{b^2}+3\frac{ b''}{b}-3\frac{H^2}{b^2}=
\frac{1}{2}\Phi'^I\Phi'_I+V(\Phi^I),
\\
&&ij: 3\frac{b'^2}{b^2}+3\frac{ b''}{b}-\frac{3H^2+2\dot{H}}{b^2}=
\frac{1}{2}\Phi'^I\Phi'_I+V(\Phi^I),
\\
&&55: 6\frac{H^2}{b^2}+3\frac{\dot{H}}{b^2}-6\frac{b'^2}{b^2}=
\frac{1}{2}\Phi'^I\Phi'_I-V(\Phi^I),
\ea
where  $H=\dot{a}/a$ is the Hubble parameter. In this paper prime denotes $y$ derivative and dot denotes time derivative. It is straightforward to observe that the $00$ and $ij$  equations lead to $\dot{H}=0$, i.e., $H$ must be a constant. Thus, we conclude that the scalar fields should be static and have $y$-dependence only.  One should note that this result is a consequence of the metric assumption (\ref{metricassumption}). For more general metric assumptions, for example, $b(y)\rightarrow b(t,y)$ and $a(t)\rightarrow a(t,y)$, we would have more general solutions with $\dot{H}\neq 0$.

We emphasize that the solution to the above equations is not unique. This is because  the Einstein equations and the scalar field equation are not independent. Due to the conservation of the energy momentum tensor, there are only three independent equations, with five indeterminate functions. Hence the system has infinite set of solutions and we are allowed to assume the solutions of two functions in advance.

\begin{figure}[htb]
  \centering
  \includegraphics[width=7.5cm]{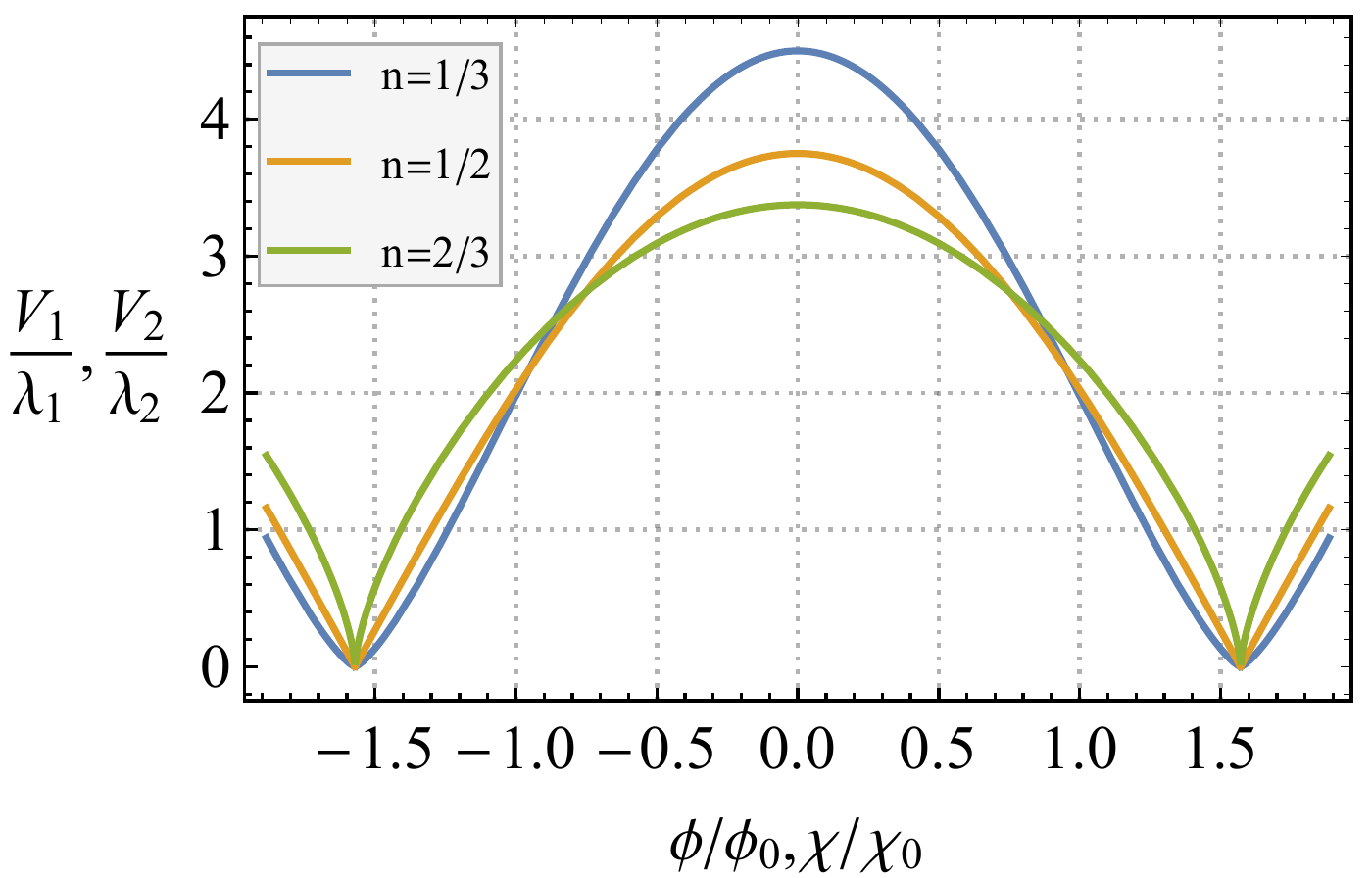}
  \includegraphics[width=8cm]{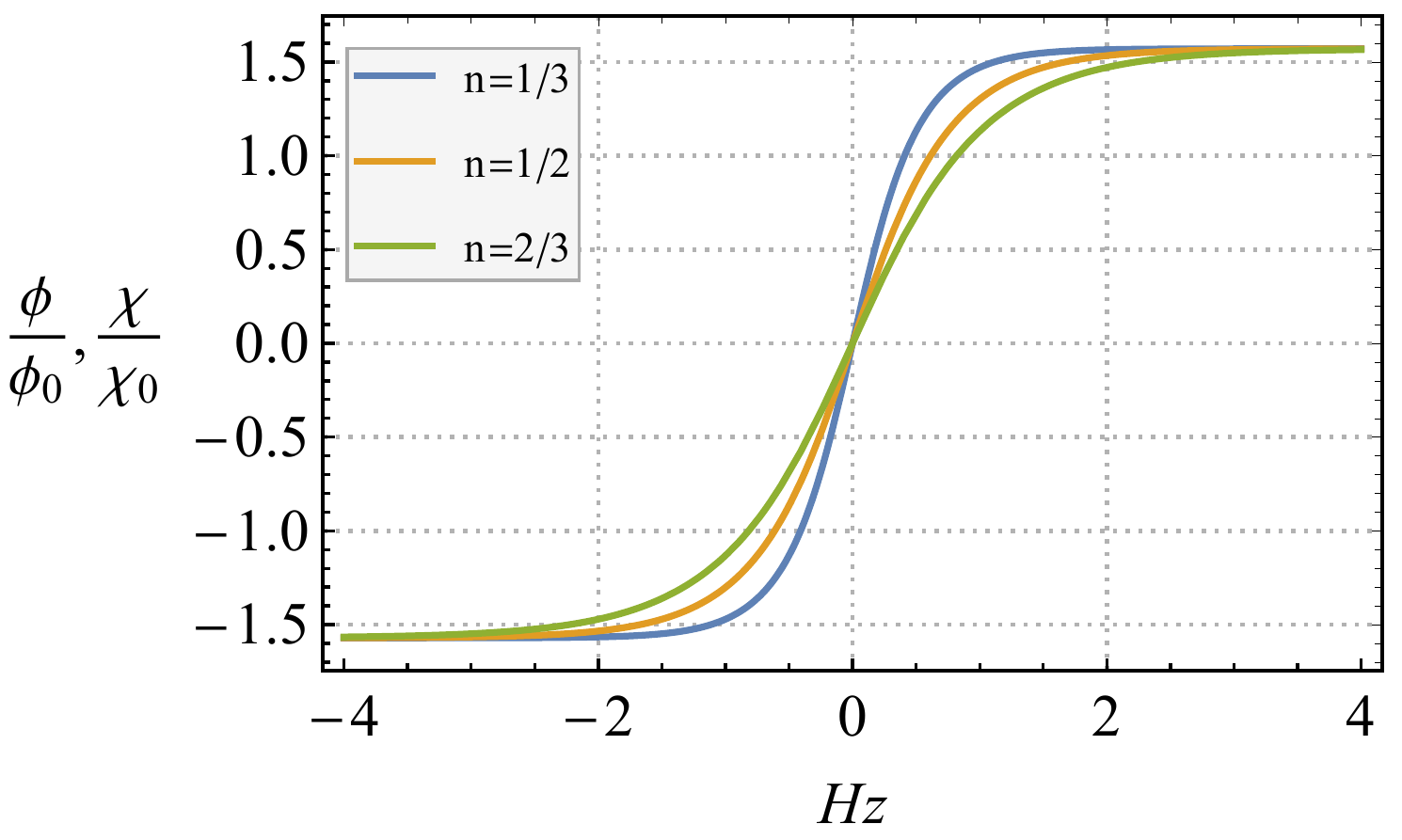}
  \caption{The scalar potentials (left) and the scalar fields (right) for different values of $n$. Note that the $|z|\rightarrow\infty$ limit corresponds to the minima of the potentials.}
  \label{potPlot}
\end{figure}
To give exact solutions, we consider a two-field theory as a concrete example. We set $\Phi_I=(\phi, \chi)$. The potential generally contains the coupling of the two fields. However, in this work we consider a decoupled example for simplicity, hence $V(\Phi_I)=V_1(\phi)+V_2(\chi)$. Under these assumptions, we give the following metric solution, which is more compact in $z$ coordinate by defining $\mathrm{d}y=b\mathrm{d}z$,
\ba
a(t)&=&\text{e}^{Ht},\\
b(z)&=&\text{sech}^{n}\left(\frac{Hz}{n}\right).
\ea
The solutions of the scalar fields are
\ba
\phi(z)&=&\phi_0\text{arctan}\left(\text{sinh}\left(\frac{Hz}{n}\right)\right),\\
\chi(z)&=&\chi_0\text{arctan}\left(\text{sinh}\left(\frac{Hz}{n}\right)\right).
\ea
The potential can also be given as
\ba
V_1(\phi)=\lambda_1\text{sec}^{2(n-1)}\left(\frac{\phi}{\phi_0}\right),\\
V_2(\chi)=\lambda_2\text{sec}^{2(n-1)}\left(\frac{\chi}{\chi_0}\right),
\ea
where the parameters are
\ba
\chi_0&=&\sqrt{\frac{3(1-n)n-\kappa_5\phi_0^2}{\kappa_5}},\\
\lambda_1&=&-\frac{3n+1}{2(n-1)n^2\kappa_5}H^2\phi_0^2, \\ \lambda_2&=&\frac{(3n+1)(3(n-1)n+\kappa\phi_0^2)}{2(n-1)n^2\kappa_5}H^2\phi_0^2.
\ea
The parameter $n$ is constrained to be $0<n<1$, and $-\sqrt{\frac{3n(1-n)}{\kappa_5}}<\phi_0<\sqrt{\frac{3n(1-n)}{\kappa_5}}$.
The plots of these solutions are shown in Figure \ref{potPlot}. It is straightforward to see that the potential has a domain wall configuration, and $|z|\sim \infty$  corresponds to its vacuum.

\section{The gravitational waves}\label{section III}
To study the gravitational waves, we consider the perturbed metric
\be
\mathrm{d}s^2=-b^2(z)\left(\mathrm{d}t^2+a^2(t)(\delta_{ij}+h_{ij})\mathrm{d}x^i\mathrm{d}x^j+\mathrm{d}z^2\right),
\ee
where $h_{ij}$ represents the tensor perturbations and thus the gravitational waves. It satisfies the transverse and traceless condition, i.e., $\delta^{ij}h_{ij}=0$ and $\partial^{i}h_{ij}=0$. In momentum space, the equation of $h_{ij}$ is
\be
\ddot{h}_{ij}+3H\dot{h}_{ij}+\frac{k^2}{a^2}h_{ij}=
3\frac{\partial_z b}{b}\partial_z h_{ij}+\partial_z^2h_{ij}.
\ee
Now using the separation of variables \cite{Langlois:2000ns},
\be
h_{ij}(t,k,z)=\int\mathrm{d}m\psi(z,m)\zeta(t,\mathbf{k},m)\hat{e}_{ij},
\ee
we can separate the equation of $h_{ij}$ into two equations,
\ba
\ddot{\zeta}_k+3H\dot{\zeta}_k+\left(\frac{k^2}{a^2}+m^2\right)\zeta_k=0,\label{cosmoPertEq}\\
\partial_z^2\psi+3\frac{\partial_z b}{b}\partial_z\psi+m^2\psi=0,\label{ShrodEq}
\ea
where the parameter $m$ is interpreted as the mass of graviton observed in four dimensions.
Equation (\ref{cosmoPertEq}) is the usual evolution equation of the tensor perturbation.  The equation (\ref{ShrodEq}) describes the component of the perturbation along the extra dimension, and this equation gives the mass spectrum of the graviton.

\subsection{The mass spectrum of the gravitons}
To study the mass spectrum of the graviton, we first transform the equation (\ref{ShrodEq}) to be a Schr\"{o}dinger-like formalism, which can be obtained by defining
\be
\Psi=b^{3/2}\psi.
\ee
Then we get the Schr\"{o}dinger-like equation of the new variable $\Psi$,
\be
-\partial_{z}^2\Psi+\left[\frac{3}{4}\left(\frac{\partial_zb}{b}\right)^2
+\frac{3}{2}\frac{\partial_z^2b}{b}\right]\Psi=m^2\Psi.
\ee
It is clear that the eigenvalue $m^2$ is completely determined by the effective potential. A very important property of this equation is that it can be factorized so that
\be
\left(\partial_z+\frac{3\partial_zb}{2b}\right)
\left(-\partial_z+\frac{3\partial_zb}{2b}\right)\Psi=m^2\Psi.
\label{factEQ}
\ee
Such a formalism ensures that $m^2\geq 0$.

\begin{figure}[htb]
  \centering
  \includegraphics[width=7.5cm]{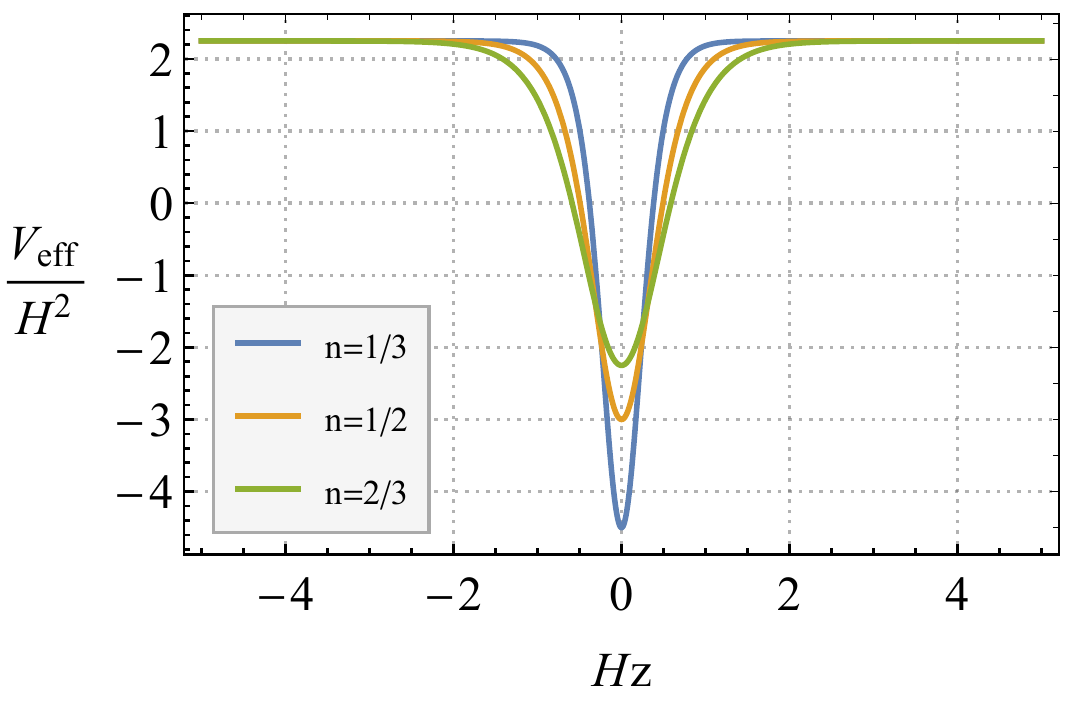}
  \includegraphics[width=7.2cm]{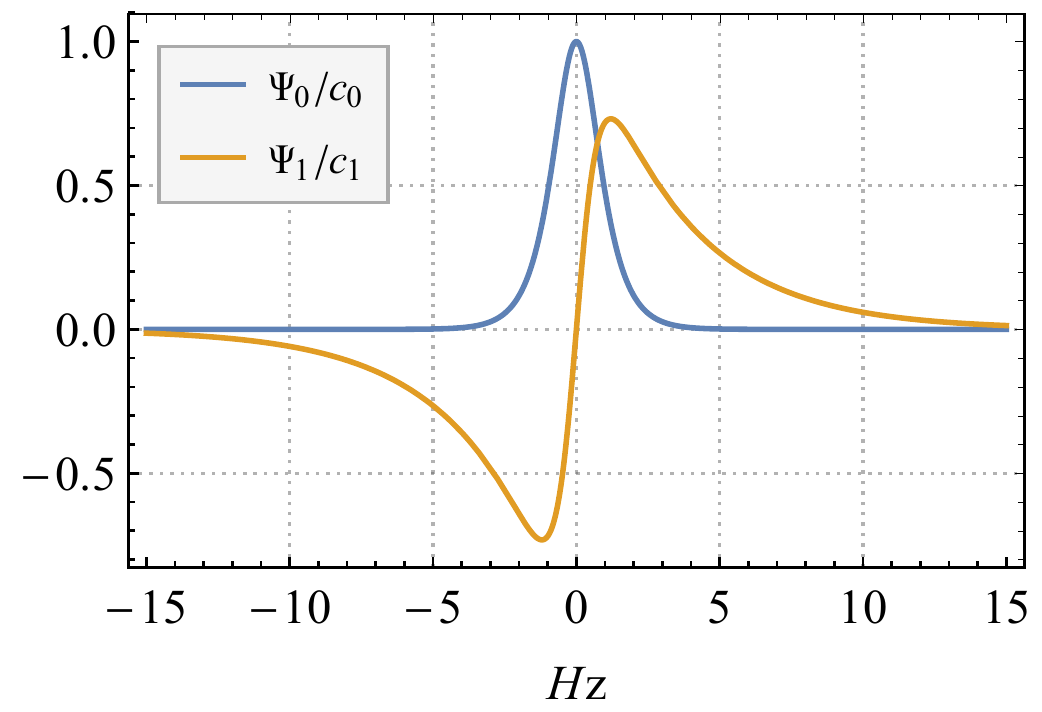}
  \caption{Left: The effective potential for different values of $n$. Right: The ground state and the first excited state with $n=5/6$.}
  \label{effPot}
\end{figure}

According to the expression of $b(z)$, we get the effective potential
\be
V_\text{eff}(z)=\frac{9H^2}{4}
\left[1-\left(1+\frac{2}{3n}\right)\text{sech}^2\left(\frac{Hz}{n}\right)\right].
\ee
The plot of this $V_{\text{eff}}$ is shown in Figure \ref{effPot}. It approaches $9H^2/4$ in the $|z|\rightarrow\infty$ limit for any value of $n$. The minimum is $V_\text{eff}(0)=-3H^2/(2n)$, which is always negative. These features imply the existence of the ground state ($m=0$) and the free states with a continuous mass spectrum ($m\geq3H/2$), which is separated by a mass gap $3H/2$. These free states are asymptotically plane waves, which can not be normalized, hence we are not interested in them.  From the equation (\ref{factEQ}), it is straightforward to get the solution of the ground state,
\be
\Psi_0(z)=c_0\text{sech}^{\frac{3n}{2}}\left(\frac{Hz}{n}\right),
\label{groundstate}
\ee
where the parameter $c_0$ can be fixed by the normalization condition. This state represents the massless graviton observed in four-dimensional spacetime.

Now the question is whether there are bound states in the mass gap $0<m<3H/2$.
To analyze the number of bound states, we write the Schr\"{o}dinger-like equation as \cite{Guo:2010az}
\be
\left[-\partial_z^2-\alpha(\alpha-1)\beta^2
\text{sech}^2\left(\frac{Hz}{n}\right)\right]\Psi=E_N\Psi,
\ee
where $\alpha=1+3n/2$, $\beta=H/n$, and $E_N=m^2-9H^2/4$. In such a formalism, the Schr\"{o}dinger-like equation can be analytically solved, giving the eigenvalues of the bound states
\be
E_N=\frac{N(3n-N)}{n^2}H^2-\frac{9}{4}H^2,\quad m=\sqrt{\frac{N(3n-N)}{n^2}}H.
\ee
The integer $N$ is constrained bo be $0\leq N \leq 3n/2$. For $0<n<2/3$, the only one possible solution is  $N=0$, which corresponds to the ground state (\ref{groundstate}). For $2/3\leq n<1$ one has $0\leq N \leq 3/2$, hence the possible solutions are $N=0$ and $N=1$, corresponding to the ground state (\ref{groundstate}) and the first excited state,
\be
\Psi_1(z)=c_1 \text{sech}^{\frac{3n}{2}}\left(\frac{Hz}{n}\right)
\sinh\left(\frac{Hz}{n}\right).
\label{FirstES}
\ee
 The mass of this state is $m_1=\sqrt{3n-1}H/n$.
Thus we conclude that there could be at most one bound state in the mass gap $0<m<3H/2$. The plots of the ground sate and the first excited state are shown in Figure \ref{effPot}. The massive graviton modes especially the bound state, would modify the gravitation at small scales, leading to corrections to the inverse square law of gravitation. The massive bound state would also modify the Standard Model of particle physics. These effects provide new approaches to detect the extra dimension.

\subsection{The cosmological evolution of gravitational waves}
Now we turn our attention to the cosmological evolution equation (\ref{cosmoPertEq}). To solve the perturbation $\zeta(t,\mathbf{k},m)$, we define a new variable $\upsilon=a\zeta$ and introduce the conformal time coordinate $\tau$ defined by $\mathrm{d}t=a\mathrm{d}\tau$. In terms of the new variable $\upsilon(\tau,\mathbf{k},m)$ and conformal time $\tau$, we get the modified Mukhanov-Sasaki equation
\be
\partial_\tau^2\upsilon_m+\left(k^2+a^2m^2-\frac{\partial^2_\tau a}{a}\right)\upsilon_m=0.
\ee
For the de Sitter solution one has $a=-1/(H\tau)$, and the above equation becomes
\be
\partial_\tau^2\upsilon_m+\left(k^2+\frac{m^2}{H^2\tau^2}-\frac{2}{\tau^2}\right)\upsilon_m=0.
\ee
The analytic solution is
\be
\upsilon_m=c_3 \sqrt{-k\tau}J_\nu(k\tau)+c_4\sqrt{-k\tau} Y_\nu(k\tau),
\ee
with $\nu=\sqrt{\frac{9}{4}-\frac{m^2}{H^2}}$. $J_\nu(k\tau)$ and $Y_\nu(k\tau)$ are the Bessel functions of the first and second kind, respectively. The behaviors of different modes at sub-hubble scales ($k\gg aH$) and super-hubble scales ($k\ll aH$) are summarized in  Table \ref{tab1}. In particular, the massless mode (corresponding to $\nu=3/2$) simply oscillates inside the horizon, and approaches a constant after it exits the horizon, the same as that of general relativity in four dimensions. However, due to the existence of extra dimension, the evolutions of gravitational waves corresponding to the massive modes are very different from that of general relativity in four dimensions. The heavy modes ($m> 3H/2$) oscillate with nearly constant amplitude when they are inside the horizon. After horizon exit, they still oscillate but with decaying amplitude, roughly $|\upsilon_m|\sim a^{-\frac{3}{2}}$.
For the light modes ($m< 3H/2$), they oscillate inside the horizon. After horizon exit they simply decay with non-oscillating behaviour. The results are summarized in Table \ref{tab1}.
\begin{table}
\caption{The evolution of $\upsilon_m$.}
\centering
\begin{tabular}{|c|c|c|}
  \hline
  ~ & $m<\frac{3}{2}H$ & $m>\frac{3}{2}H$ \\
  \hline
  $k\gg aH$ & oscillating, $|\upsilon_m| \sim$ constant & oscillating, $|\upsilon_m| \sim$ constant  \\
  \hline
  $k\ll aH$ & non-oscillating, $|\upsilon_m|\sim a^{\nu-\frac{3}{2}}$ & oscillating, $|\upsilon_m| \sim a^{-\frac{3}{2}}$ \\
  \hline
\end{tabular}
\label{tab1}
\end{table}

\section{Conclusions}\label{concs}
In this paper we constructed a braneworld model with multi scalar fields. The extra dimension is infinitely large and the fields are allowed to be in the five dimensional bulk. We considered a four dimensional de Sitter solution as an approximation to the early universe inflation. The scalar fields have kink solutions. The scalar potential has a domain wall configuration. Under such a background solution, we studied the tensor perturbations. We showed that the graviton has a series of massive Kaluza-Klein modes. These massive modes are separated from the massless mode by a $3H/2$ mass gap. There could be a massive bound state in this mass gap, depending on the background solutions.
We also studied the cosmological evolution of the gravitational waves. All the modes oscillate when they are sub-horizon scales. However, after they exit the horizon the light modes simply decay with non-oscillating behaviour,  and the heavy modes still oscillate but with decaying amplitude. These features distinguish the extra dimension theory from the conventional four dimensional theory, and could potentially be used to detect extra dimension.

\section{Acknowledgements}
This work was supported in part by the National Natural Science Foundation of China (Grants No. 11947025 and No. 12165013) and by the Innovation Training Program for College Students of Jiangxi province.

\end{document}